\documentstyle[12pt,epsfig]{article}
\begin{document}
\newcommand{\beq}{\begin{equation}}
\newcommand{\eeq}{\end{equation}}
\newcommand{\barr}{\begin{eqnarray}}
\newcommand{\earr}{\end{eqnarray}}

\newcommand{\andy}[1]{ }

\def\txt{\textstyle}
\def\ask{\marginpar{?? ask:  \hfill}}
\def\fin{\marginpar{fill in ... \hfill}}
\def\note{\marginpar{note \hfill}}
\def\check{\marginpar{check \hfill}}
\def\discuss{\marginpar{discuss \hfill}}
\def\hh{\widehat}
\def\tt{\widetilde}
\def\cH{{\cal H}}
\def\cT{{\cal T}}
\def\cR{{\cal R}}
\def\cC{{\cal C}}
\def\cZ{{\cal Z}}
\def\cL{{\cal L}}
\def\txt{\textstyle}
\def\bmp{\mbox{\boldmath $p$}}
\def\bmk{\mbox{\boldmath $k$}}
\def\bmksub{\mbox{\boldmath\scriptsize $k$}}
\def\bmp{\mbox{\boldmath $p$}}
\def\bmr{\mbox{\boldmath $r$}}
\def\bmA{\mbox{\boldmath $A$}}
\def\bmv{\mbox{\boldmath $v$}}
\def\bmg{\mbox{\boldmath $g$}}
\def\bmepsilon{\mbox{\boldmath $\epsilon$}}
\def\bmhA{\mbox{\boldmath $\hat{A}$}}
\def\bmhp{\mbox{\boldmath $\hat{p}$}}
\def\bmhv{\mbox{\boldmath $\hat{v}$}}
\def\pade#1#2{{frac{\partial#1}{\partial#2}}}
\newcommand{\mean}[1]{\langle #1 \rangle}

\begin{titlepage}
\begin{flushright}
\today \\
\end{flushright}
\vspace{.5cm}
\begin{center}
{\LARGE Classical and quantum dynamics of a particle constrained
on a circle}

\quad

{\large Antonello Scardicchio \footnote{email:Antonello.Scardicchio@ba.infn.it}\\
\quad
\\}

{\large Dipartimento di Fisica, Universit\`a di Bari \\ \quad \\
 I-70126  Bari, Italy}

\vspace*{.5cm} PACS: 04.60.Ds, 03.65.Db\vspace*{.5cm}

{\small\bf Abstract}\\ \end{center}

{\small

The Dirac method is used to analyze the classical and quantum
dynamics of a particle constrained on a circle. The method of
Lagrange multipliers is scrutinized, in particular in relation to
the quantization procedure. Ordering problems are tackled and
solved by requiring the hermiticity of some operators. The
presence of an additional term in the quantum Hamiltonian is
discussed.

}

\end{titlepage}

\newpage
\setcounter{equation}{0}
\section{Introduction}
\label{sec-intro}
\andy{sec-intro}

The seminal and, so far, most used way to formulate the quantum
theory of a particle or a field makes wide use of the Hamiltonian
description of classical mechanics \cite{DiracBook}. The standard
rules for constructing the momenta and the Hamiltonian function,
however, cannot be applied when the Lagrangian is singular. In
such a case it is not possible to extract the functional
dependence of all the velocities on the momenta in order to obtain
a Hamiltonian function of coordinates and momenta only. Dirac's
method concerns the study of classical systems using the
Hamiltonian method when the usual procedure fails due to the
singularity of the Lagrangian \cite{Dirac}. Dirac gave very
general rules to construct the Hamiltonian and calculate sensible
brackets that can be used to describe the classical and, by the
canonical quantization procedure, the quantum dynamics.

One of the most interesting situation where Dirac's method of
handling singular Lagrangians can be applied is in confining
particles on curved manifolds \cite{Regge,Kleinert}. Part of this
interest is certainly due to the presence of additional terms
which arise in many quantization procedures on curved manifolds
\cite{DeWitt,SchulmanBook,Schulman} and is far from being
clarified. In this letter we will focus our attention on the
connection between the additional terms which occur in the quantum
Hamiltonian and the problem of the operator ordering prescription.

In Section 2 we briefly review Dirac's method of handling singular
Lagrangians. In Section 3 we quantize a free particle constrained
on a circle following the standard procedure, i.e.\ reducing from
the very beginning the number of degrees of freedom. Then we solve
the same (classical) problem using Dirac's method, recovering a
new set of canonical brackets. Finally we quantize using this
bracket algebra, by focusing our attention on the construction of
coordinates, linear momenta, angular momentum and Hamiltonian
operators and on related ordering problems and we will finally
write the Schrodinger equation. Section 4 contains our
conclusions.
\section{The Dirac method}
\label{sec-diracQ}
\andy{sec-diracQ}
Let us start by outlining the Dirac method \cite{Dirac} and
introduce notation. Take a consistent Lagrangian $L(x,\dot{x})$
with $N$ coordinates. The classical dynamics is obtained by the
least action principle:
\barr
S[x]=\int_{t_0}^{t_1}dt L(x,\dot{x}), \nonumber \\
\delta S[x]=0, \earr which in terms of the Lagrangian gives $N$
Euler-Lagrange equations
\beq
\frac{\partial L}{\partial x_i}-\frac{d}{dt}\frac{\partial
L}{\partial \dot{x}_i}=0.
\eeq
We define momenta and Hamiltonian and obtain the usual (Poisson)
brackets between momenta and coordinates:
\andy{eq:momenta}
\barr
p_i &=& \frac{\partial L}{\partial \dot x_i}, \qquad (i=1,...,N)
\label{eq:momenta1} \\
H(x,p) &=& \sum_i p_i \dot x_i(x,p)-L(x,\dot x(x,p)), \label{eq:momenta2} \\
\left[ x_i, p_j \right] &=& \delta_{ij}, \label{eq:momenta3} \earr
and for any function $A$ of $x$'s and $p$'s (not explicitly
dependent on time),
\beq
\dot A=[A,H].
\label{eq:momenta4}
\eeq
Two scenarios are possible. In the typical case one can invert
$p_i(x,\dot{x})$ to obtain $\dot{x}_i(x,p)$; if this is not
possible, not even locally, the Lagrangian is said singular and
its Hessian with respect to the velocities vanishes
\beq \left
|\left|\frac{\partial L}{\partial \dot x_i \partial \dot
x_j}\right|\right|=0.
\eeq
In such a case we act differently. We consider those relations in
(\ref{eq:momenta1}) which hinder the inversion (this step will be
clarified in the example of Section 3) as a series of
\emph{constraints}
\beq
\phi_j\approx 0
\eeq
which must be satisfied ``weakly" (namely, their Poisson bracket
with any given quantity may not vanish) along the physical
trajectory. In this way we obtain a number (say $M$) of
constraints which Dirac called \emph{primary} because of their
direct derivation from the Lagrangian. Notice that a Hamiltonian
is required to be independent of the velocities. If we are not
able to erase the $\dot{x}$ dependence, then the straightforward
application of the hamiltonian method is impossible. To solve our
problem we proceed as follows. We add to $H$ all our primary
constraints multiplied by arbitrary functions of time $u_j$, to
obtain the total Hamiltonian $H_T$
\andy{eq:totalhamilt}
\beq
H_T=H+\sum^{M}_{j=1} u_j \phi_j(x,p).
\eeq
This could seem to imply an arbitrariness (additional freedoms are
introduced) but we require a number of consistency conditions:
each constraint must be zero during the whole evolution, if it is
initially zero:
\andy{eq:consistency}
\beq
\dot{\phi_j}=[\phi_j,H_T]\approx 0 \qquad (j=1,...,M).
\label{eq:consistency}
\eeq
If these equations are consistent, three cases are possible: an
equation can give an identity; it can give a linear equation for
the $u_j$; it can give an equation containing only $p$'s and
$x$'s, in which case it must be considered as another constraint.
The constraints that arise from this procedure will be called
\emph{secondary}, for obvious reasons. Even for these, we impose
consistency conditions and this procedure is continued until we
have a set of identities and linear equations for the $u$'s. Now
we have enlarged our set of constraints to include the secondary
ones and we have a new number of constraints, say $K$.

We have by now defined a constraint as a quantity which satisfies
\andy{eq:consistency2,3}
\barr
\label{eq:consistency3}
\phi_j &\approx &0, \\
\label{eq:consistency2}
[\phi_j,H_T]&\approx &0. \earr This defines a linear vector space
(due to the linearity of the Poisson brackets) and so any linear
combination of constraints is again a constraint. It is of great
importance for our purposes the distinction between \emph{first
class} and \emph{second class} constraints. The first are defined
as the constraints which ``commute" (i.e.\ have vanishing Poisson
brackets) with all the other constraints. The second ones have at
least one non vanishing bracket with some other constraint. It may
happen that we can take linear combinations of second class
constraints and obtain some first class constraints. This
situation brings to light the presence of some gauge degrees of
freedom. Dirac showed the profound difference between this two
classes. In fact we can switch to new canonical brackets in order
to set all of our \emph{second} class constraints \emph{strongly}
equal to zero. This means that in any given quantity, such as the
Hamiltonian, we can set them to zero ``by hand". The first class
ones, however, will ``survive" (even in the Hamiltonian with their
arbitrary multiplicative functions $u$). In the following analysis
we will not deal with first class constraints and so will not
discuss them any further. Every constraint that we will find will
be of the second class. In such a case, we can safely change to
the new canonical brackets, the so called \emph{Dirac brackets},
defined as follows: let
\beq
M_{ij}\equiv [\phi_i,\phi_j]  \nonumber
\eeq
and its inverse
\beq
G_{ij}\equiv (M^{-1})_{ij} \nonumber
\eeq
(the invertibility of $M$ is a particular feature of the absence
of first class constraints: in general $M$ is defined on the
subspace of second class constraints only). Then for any two
quantities $A$ and $B$ we define the Dirac bracket:
\andy{eq:diracB}
\beq
[A,B]_D=[A,B]-\sum^K_{i,j=1}[A,\phi_i]G_{ij}[\phi_j,B].
\label{eq:diracB}
\eeq
These brackets have all the properties of the Poisson bracket plus one: for
any dynamical variable $A$ we have
\andy{eq:strongzero}
\barr
[A,\phi_i]_D=0,
\label{eq:strongzero1} \\
\dot{A}=[A,H_T]\approx [A,H_T]_D,
\label{eq:strongzero2}
\earr as is easy to see (for (\ref{eq:strongzero2}) use
(\ref{eq:consistency2})).

The very meaning of this redefinition of the canonical brackets is
simply a change of variables from the original phase space to the
constrained manyfold \cite{Regge}. Having obtained a set of
canonical brackets, we can now quantize, by looking for
self-adjoint operators which satisfy the canonical commutation
relation (each quantity in the righthand side must be multiplied
by $i\hbar$).

Let us now look at an interesting example.

\section{Particle on a circle}
\setcounter{equation}{0}
\label{sec:particleoncircle}
\andy{sec:particleoncircle}
\subsection{The standard approach}
We want to quantize the following free particle Lagrangian
\andy{eq:freelagr}
\beq
\label{eq:freelagr}
L=\frac{1}{2}\left(\dot{x}^2+\dot{y}^2\right),
\eeq
subject to the relation
\andy{eq:circleconstr}
\beq
\label{eq:circleconstr}
r^2\equiv x^2+y^2=r_0^2
\eeq
($r_0$ being a positive real constant) which must be satisfied at
any time. This describes the motion of a particle of unitary mass
in the $xy$-plane, constrained on a circle of radius $r_0$. We can
make a change of variables, from cartesian to polar coordinates
$(r,\theta)$,
\barr
x&=&r\cos\theta, \\
y&=&r\sin\theta, \nonumber \earr after which, using
(\ref{eq:circleconstr}), the Lagrangian reads
\beq
L=\frac{1}{2}r_0^2\dot{\theta}^2.
\eeq
We have now a new Lagrangian with only one degree of freedom
$\theta$. We can define the momentum $p_\theta$
\beq
p_\theta=\frac{\partial L}{\partial \dot\theta}=r_0^2\dot{\theta}
\eeq
and the Hamiltonian
\beq
\label{eq:standclasshamilt}
H(\theta,p_\theta)=\dot{\theta}p_\theta-L=\frac{p_\theta
^2}{2r_0^2}.
\eeq
The radial degree of freedom $r$ disappears (as
implicitly did any other non-dynamical degree of freedom, such as
the $z$ coordinate in (\ref{eq:freelagr})). The Poisson bracket is
\beq
[\theta,p_\theta]=1.
\eeq

Now, let us quantize: define two self-adjoint operators
$\hat{\theta}$ and $\hat{p}_\theta$ satisfying the canonical
commutation relation (CCR) ($\hbar$=1):
\andy{eq:CCRtheta}
\beq
\label{eq:CCRtheta}
[\hat{\theta},\hat{p}_\theta]=i
\eeq
(we shall use the same notation for Poisson brackets and
commutator of operators, since no confusion can arise). We can
find such a couple of self-adjoint operators in the Hilbert space
$\cH=L^2(0,2\pi)$ and their expression is:
\barr
\label{eq:standthetap}
\hat{\theta}\psi(\theta)&=&\theta\psi(\theta),\\
\hat{p}_\theta\psi(\theta)&=&\left(-i\frac{\partial}{\partial
\theta}-\alpha\right)\psi(\theta). \nonumber
\earr
We add the constant $\alpha$ in the momentum $p_\theta$ to mimic
the possible presence of a magnetic field enclosed in the circle
(see the discussion after (\ref{eq:anticommformP})). Their domains
are chosen to be respectively $D_\theta=\cH$ and
$D_{p_\theta}=\{\psi\in\cH|\psi(0)=\psi(2\pi),\psi'\in \cH\}$.
These are dense subsets of $\cH$. Notice also that we have chosen
one of the infinite self-adjoint extensions of the momentum
$\hat{p}_\theta$. The Hamiltonian reads \andy{eq:standhamilt} \beq
\label{eq:standhamilt}
H(\hat{\theta},\hat{p}_\theta)=\frac{\hat{p}_\theta
^2}{2r_0^2}=\frac{1}{2r_0^2}\left(-i\frac{\partial}{\partial
\theta}-\alpha\right)^2, \eeq and is self-adjoint in the domain of
$p_\theta$, i.e.\ $D_{p_\theta}$. The Schrodinger equation is
(reinserting $m$ and $\hbar$) \beq \label{eq:standardschreq}
i\hbar\frac{\partial \psi}{\partial
t}=\frac{\hbar^2}{2mr_0^2}\left(
-i\frac{\partial}{\partial\theta}-\alpha \right)^2\psi. \eeq This
is what we expected.

\subsection{Dirac's approach}

Let analyze the same problem with Dirac's method. We start from
classical dynamics. We want to find the extremum of the action
with the Lagrangian defined in (\ref{eq:freelagr}), subject to the
constraint
\beq
\label{eq:phiconstr}
\phi=x^2+y^2-r_0^2 \approx 0.
\eeq
We use the method of Lagrange multipliers \cite{Elsgolts} and
search for the extremum of the action with the new Lagrangian
\andy{eq:constrlagrang}
\beq
\label{eq:constrlagrang}
L(x,\dot{x},y,\dot{y},\lambda)=\frac{1}{2}\dot{x}^2+\frac{1}{2}\dot{y}^2-\lambda(x^2+y^2-r_0^2),
\eeq
the quantity $\lambda$ being treated as an additional dynamical
variable. This Lagrangian gives rise to an action functional
$S[x,y,\lambda]$ which must be varied with respect to $x, y$ and
also the ``new" degree of freedom $\lambda$. If we want to use the
Hamiltonian method with this Lagrangian, we must start by
calculating the momenta:
\andy{eq:constrmomenta}
\barr
\label{eq:constrmomenta}
p_x&=&\frac{\partial L}{\partial \dot{x}}=\dot{x}, \nonumber \\
p_y&=&\frac{\partial L}{\partial \dot{y}}=\dot{y}, \\
p_\lambda &=&\frac{\partial L}{\partial \dot{\lambda}}=0.
\nonumber \earr

It is apparent that we are facing the situation discussed in
the Introduction and in Sec.\ 2: one of the momenta disappears. So
we proceed as previously sketched: read the relation
$p_\lambda\approx 0$ as a primary constraint:
\andy{eq:firstphi}
\beq
\label{eq:firstphi}
\phi_1=p_\lambda \approx 0.
\eeq
This is our only primary constraint. Build up the Hamiltonian
\andy{eq:constrhamilt}
\beq
\label{eq:constrhamilt}
H=p_x \dot{x}+p_y
\dot{y}+p_\lambda\dot{\lambda}-L=\frac{p_x^2}{2}+\frac{p_y^2}{2}+p_\lambda\dot{\lambda}+\lambda
(x^2+y^2-r_0^2).
\eeq
We now include $\phi_1$ multiplied by an arbitrary function of the
time $u_1$:
\andy{eq:constrtothamilt}
\beq
\label{eq:constrtothamilt}
H_T=\frac{p_x^2}{2}+\frac{p_y^2}{2}+\lambda
(x^2+y^2-r_0^2)+u_1 p_\lambda.
\eeq
Notice that $\dot{\lambda}$ has been absorbed in the arbitrary
function $u_1$. The consistency condition (\ref{eq:consistency})
is
\beq
0\approx
\dot{\phi}_1=[\phi_1,H_T]=[p_\lambda,H_T]=-\left(x^2+y^2-r_0^2\right),
\eeq
which is a new constraint, that the Lagrange multipliers
had already implicitly imposed ($\phi$ in (\ref{eq:phiconstr}))
\beq
\phi_2=\phi=x^2+y^2-r_0^2\approx 0.
\eeq
The consistency conditions (\ref{eq:consistency}) for $\phi_2$
yields
\beq
\label{eq:constrphi3}
\phi_3=xp_x+yp_y\approx 0
\eeq
and by imposing (\ref{eq:consistency}) also for $\phi_3$ we obtain
\beq
\phi_4=p_x^2+p_y^2-2(x^2+y^2)\lambda\approx 0.
\eeq
These are additional constraints. If we impose
(\ref{eq:consistency}) for $\phi_4$ we get an equation for $u_1$:
\beq
u_1=-\frac{2 \lambda}{x^2+y^2}(xp_x+yp_y)\approx 0.
\eeq

Since in the following we shall use only Dirac brackets we regard
any constraint as a \emph{strong} equation and drop the term $u_1
\phi_1$ from the total Hamiltonian. We can also drop the term
containing the Lagrangian multiplier because of $\phi_2$. So our
Hamiltonian becomes the free one:
\beq
\label{eq:diarctothamilt}
H_T=\frac{p_x^2}{2}+\frac{p_y^2}{2}.
\eeq The fact that the
Hamiltonian function of the constrained dynamics is exactly that
of an unconstrained dynamics may seem strange. One could
(erroneously) argue that even the equations of motion would be the
same. This is not correct because we will change the canonical
brackets. All additional information characterising the
constrained dynamics is now contained in these new canonical
brackets. One could say that Dirac's method ``drains" information
from the Lagrangian, where it is contained in the additional
degree of freedom $\lambda$, giving it to the canonical brackets,
where it is contained in a non-trivial algebra. In this process,
however, the information on the topology of the problem is made
explicit, as we shall see in the short discussion just after the
algebra construction. This point of view is very useful in quantum
mechanics.

We have four constraints and what we need now is the algebra of
the Dirac's brackets. We calculate the matrix ($\bmr=(x,y)$ and
$\bmp=(p_x,p_y)$) \beq M=\begin{tabular}{|rrrr|}
         0 &          0 &          0 &          $2\bmr^2$ \\
         0 &          0 &          $2\bmr^2$ &          $4\bmp\cdot \bmr$\\
         0 &         $-2\bmr^2$ &          0 &          $2\bmp^2+4\lambda \bmr^2$ \\
        $-2\bmr^2$ &    $-4\bmp\cdot \bmr$ &         $-2\bmp^2-4\lambda \bmr^2$ &          0 \\
\end{tabular}
\eeq
and invert it to get
\beq
G=
\begin{tabular}{|rrrr|}
         0 &            $-(\bmp^2+2\lambda \bmr^2)/2\bmr^4$ &   $\bmr\cdot\bmp/\bmr^4$ &  $-1/2\bmr^2$ \\
         $(\bmp^2+2\lambda \bmr^2)/2\bmr^4$ &             0 & $-1/2\bmr^2$ &               0 \\
         $-\bmr\cdot\bmp/\bmr^4$ & $1/2\bmr^2$ &             0 &               0 \\
$1/2\bmr^2$ &          0 &             0 &               0 \\
\end{tabular}\quad .
\eeq
We can now calculate the Dirac brackets of any two quantities and
appreciate their physical meaning.

To start off, let us first consider an interesting example of the
difference between Poisson and Dirac brackets. We can check
whether (\ref{eq:strongzero1}) is true for $\phi_1=p_\lambda$ and
$A=\lambda$. The commutation rule between the Lagrange multiplier
and its momentum changes from $[\lambda,p_\lambda]=1$ to \barr
[\lambda,p_\lambda]_D&=&1-\sum_{i,j}[\lambda,\phi_i]G_{ij}[\phi_j,p_\lambda]= \nonumber\\
&=&1-[\lambda,\phi_1]G_{1
4}[\phi_4,p_\lambda]=1-1\left(-\frac{1}{2r^2}\right)(-2r^2)=0,
\nonumber
\earr which enables one to see how the Dirac brackets work in
order to satisfy the constraints strongly. We also find (we have
replaced $r$ with $r_0$ in each quantity by using $\phi_2=0$):
\andy{eq:DiracCCR}
\barr
\label{eq:DiracCCR}
\left[ x,p_x \right]_D&=&1-\frac{x^2}{r_0^2},\nonumber  \\
\left[ y,p_y \right]_D&=&1-\frac{y^2}{r_0^2}, \nonumber \\
\left[ x,p_y \right]_D&=&-\frac{x y}{r_0^2}, \\
\left[ y,p_x \right]_D&=&-\frac{x y}{r_0^2}, \nonumber \\
\left[ x,y \right]_D&=&0, \nonumber \\
\left[ p_x,p_y \right]_D&=&-\frac{1}{r_0^2}(x p_y -y p_x).
\nonumber \earr This brackets have a nice geometric
interpretation. According to the Poisson bracket $[x,p_x]=1$,
$p_x$ is the \emph{generator of translations} along the $x$ axis.
However this property cannot be preserved in the constrained
algebra, because typically we cannot translate in the $x$
direction while remaining on the circle. This can be done only at
the points $(x=0,y=\pm r_0)$ where the first and fourth equations
of (\ref{eq:DiracCCR}) reduce to the Poisson algebra. Another
feature is to be noticed: $x$ and $y$ still commute. We can
understand this because $x$ and $y$ are the generators of
translations in the corresponding $p$'s directions; however there
is no constraint containing \emph{only} momenta so any given point
in the $p_x p_y$-plane is allowed, by suitably adjusting the other
coordinates $x,y$ and $\lambda$. This is not the case of the
coordinates $x$ and $y$, as one can readily see: for example, the
point $x=2r_0,y=r_0$ is not allowed even by making additional
translations of momenta and $\lambda$, because of $\phi_2$.

We can write the Hamiltonian in the form (\ref{eq:standhamilt}) defining $L_z$:
\beq
\label{eq:classlz}
L_z=xp_y-yp_x.
\eeq
Squaring it and using $\phi_2$ we obtain
\beq
\label{eq:lzsquared}
L_z^2=r_0^2(p_x^2+p_y^2)-(xp_x+yp_y)^2,
\eeq
and using $\phi_3$ we obtain
\beq
\label{eq:classhamiltL0}
H=\frac{1}{2}(p_x^2+p_y^2)=\frac{L_z^2}{2 r_0^2}.
\eeq
One can identify $L_z$ with $p_\theta$ by writing coordinates and
momenta as functions of $\theta$ and $L_z$. This can be done by
solving the equations (\ref{eq:classlz}) and (\ref{eq:constrphi3})
for $p_x$ and $p_y$ and using (\ref{eq:circleconstr}). We get:
\barr
\label{eq:xpfunctofLtheta1}
x&=&r_0\cos\theta,\quad y=r_0\sin\theta, \\
\label{eq:xpfunctofLtheta2}
p_x&=&-\frac{1}{r_0}L_z\sin\theta,\quad
p_y=\frac{1}{r_0}L_z\cos\theta.
\earr The reader can verify that
all the relations obtained by the Dirac brackets algebra are
equivalent to the single bracket $[\theta,L_z]=1$ (e.g.\
$[x,L_z]_D=-y$ should be read $[\cos\theta,L_z]=-\sin\theta$ and
so on).

Equations (\ref{eq:DiracCCR}) pave the way to quantization. We
shall see that the quantization of the Dirac algebra is not a
trivial problem: our recipe will be the requirement that some
operators be self-adjoint (or at least Hermitian). This
requirement will play a fundamental role in our analysis. We look
at an explicit representation of the self-adjoint operators
$\hat{x},\hat{y},\hat{p}_x,\hat{p}_y$ (notice that we will not
deal with the operators $\hat{p}_\lambda$ and $\hat{\lambda}$
because they are completely defined by $\phi_1 =0$ and $\phi_4=0$
respectively) satisfying this algebra. We must, however, impose
the (now) strong equalities $\phi_i=0\quad (i=1,2,3,4)$. So (in
the following we will drop all hats on operators), $r^2\equiv
x^2+y^2=r_0^2$ and there exists a self-adjoint operator $\theta$
on the Hilbert space $\cH=L^2(0,2\pi)$ such that: \barr
\label{eq:Opxy}
x&=&r_0\cos \theta, \\
y&=&r_0\sin \theta. \nonumber \earr
We will determine the momentum operators in order to satisfy the following
equations:
\andy{eq:OpDiracCCR}
\barr
\label{eq:OpDiracCCR}
\left[ x,p_x \right]&=&i\left(1-\frac{x^2}{r_0^2}\right) \nonumber\\
\left[ y,p_y \right]&=&i\left(1-\frac{y^2}{r_0^2}\right)\nonumber \\
\left[ x,p_y \right]&=&-i\frac{x y}{r_0^2}  \\
\left[ y,p_x \right]&=&-i\frac{x y}{r_0^2} \nonumber \\
\left[ x,y \right]&=&0 \nonumber \\
\left[ p_x,p_y \right]&=&-\frac{i}{r_0^2}(x p_y -y p_x). \nonumber
\earr Using the fact that ($d_\theta$ stands for the
$\theta$-derivative, $F$ for any $n$-times differentiable
function) $[d_\theta^n,F(\theta)]$ contains derivatives of order
less than or equal to $n-1$ and looking at the first two equations
in (\ref{eq:OpDiracCCR}) (whose right hand side does not contain
momenta) one can infer that the $p$ operators in the $\theta$
representation contain only first order derivatives. Then, in the
most general case,
\barr
p_x&=&-\frac{i}{r_0} f(\theta)\frac{\partial}{\partial \theta}+\frac{1}{r_0}a(\theta),\\
p_y&=&-\frac{i}{r_0} g(\theta)\frac{\partial}{\partial
\theta}+\frac{1}{r_0}b(\theta).
\earr
Using these expressions we solve for the unknown functions $f,g,a$
and $b$. The first equation in (\ref{eq:OpDiracCCR}) yields
\beq
\cos\theta\left(-if(\theta)\frac{\partial}{\partial \theta
}\right)-\left(-if(\theta)\frac{\partial}{\partial
\theta}\right)\cos\theta=i\sin^2\theta, \nonumber
\eeq
which is solved to give
\beq f(\theta)=-\sin\theta \nonumber.
\eeq
Analogously, the solution of the second equation in
(\ref{eq:OpDiracCCR}) gives
\beq
g(\theta)=\cos\theta.
\eeq
At
this stage the third, fourth and fifth equations in
(\ref{eq:OpDiracCCR}) are identities and yield no information on
$a$ and $b$. However, some insight on their form can be obtained
from the last of (\ref{eq:OpDiracCCR}), which gives
\beq
a'\cos\theta +b'\sin\theta=-b \cos\theta+a \sin\theta,
\eeq
where the primes denotes derivatives. This yields
\barr
a'&=&-b, \\
b'&=&a. \nonumber
\label{eq:abeqn1}
\earr
However, there are other equations which must be satisfied:
\barr
\left[x,H\right]&=&i p_x, \\
\left[y,H\right]&=&i p_y. \nonumber
\earr
These are linearly dependent and both equivalent to
\beq
ia \cos\theta+ib\sin\theta=-\frac{1}{2}.
\label{eq:abeqn2}
\eeq
By using (\ref{eq:abeqn1}) this turns into an equation for $a$
whose solutions, under the additional requirement that $p_x$ and
$p_y$ be hermitian operators (this is a necessary step in order to
require their self-adjointness), are:
\barr
a(\theta)&=&\frac{i}{2}\cos\theta+\alpha\sin\theta, \nonumber \\
b(\theta)&=&-a'=\frac{i}{2}\sin\theta-\alpha\cos\theta, \nonumber
\earr
where $\alpha$ is an arbitrary \emph{real} number. Putting
all the results together we obtain
\barr
\label{eq:Opsolutforp}
p_x&=&\frac{i}{r_0}\sin\theta\frac{\partial}{\partial \theta}+\frac{i}{2r_0}\cos\theta+\frac{\alpha}{r_0}\sin\theta, \\
p_y&=&-\frac{i}{r_0}\cos\theta\frac{\partial}{\partial
\theta}+\frac{i}{2r_0}\sin\theta-\frac{\alpha}{r_0}\cos\theta.
\nonumber
\earr
We can put these equations in a compact form by using the
anticommutator (for any operators $A$ and $B$: $\{A,B\}\equiv
AB+BA$):
\barr
\label{eq:anticommformP}
p_x&=&\frac{1}{2r_0}e^{i\alpha\theta} \left\{
i\frac{\partial}{\partial
\theta},\sin\theta \right\} e^{-i\alpha\theta}, \\
p_y&=&\frac{1}{2r_0}e^{i\alpha\theta}\left\{
i\frac{\partial}{\partial \theta},-\cos\theta\right\}
e^{-i\alpha\theta}. \nonumber
\earr
Written in this form, these equations readily show some properties
of these operators. First, they are the Weyl ordered operators of
the classical quantities (\ref{eq:xpfunctofLtheta2}) but this
ordering arises naturally by taking suitable solutions of the
algebra equations. Second, these $p$'s are self-adjoint in the
domain $D_{p_\theta}$ defined after (\ref{eq:standthetap}).
Finally, equations (\ref{eq:anticommformP}) also show that
different $p$'s, corresponding to different $\alpha$'s, are
connected to each other by means of gauge transformations; this
interesting property can be easily related to the Aharonov-Bohm
effect (see \cite{Sakurai}), identifying $\alpha$ with
$\frac{e}{2\pi c}\Phi_B$ where $\Phi_B$ is the flux of the
magnetic field enclosed in the circle.

One can check that all the constraints are satisfied: remember
that we have chosen the expressions of $x$ and $y$ to satisfy
$\phi_2$, set $p_\lambda=0$ to satisfy $\phi_1$, defined $\lambda$
to satisfy $\phi_4$, so we must manage only with $\phi_3$.
Physically $\phi_3/r_0$ is the radial part of the momentum
$p_r\equiv (\bmr\cdot \bmp)/r_0$ (the vector $\bmr$ being on the
circle: $\bmr^2=r_0^2$) so we choose to represent it with a
Hermitian operator. We therefore order it ($W$ stands for `Weyl
ordering' which coincides with any other sufficiently symmetric
operator ordering procedure for this simple quantity) in order to
get the Hermitian expression
\beq
\phi_{3,W}=\frac{1}{2}\left(\{x,p_x\}+\{y,p_y\}\right).
\eeq

One can easily see, using the solutions (\ref{eq:Opxy}) and
(\ref{eq:Opsolutforp}), that $\phi_{3,W}=0$. Conversely, using the
algebra relations (\ref{eq:OpDiracCCR}) it is possible to show
that if a non Weyl-ordered expression for $\phi_3$ is constrained
to zero the momentum operators are not Hermitian. In fact using
the algebra of commutators (\ref{eq:OpDiracCCR}) one readily gets
three equivalent expressions for the momenta operators:
\barr
p_x&=&\frac{1}{2r_0^2}\left\{-y,L_z\right\}+\frac{1}{r_0^2}x\left(-\frac{i}{2}+xp_x+yp_y\right) \nonumber \\
&=&\frac{1}{2r_0^2}\left\{-y,L_z\right\}+\frac{1}{r_0^2}x\left(\frac{i}{2}+p_x
x+p_y y\right)\nonumber \\
&=&\frac{1}{2r_0^2}\left\{-y,L_z\right\}+\frac{1}{r_0^2}x\phi_{3,W}, \\
p_y&=&\frac{1}{2r_0^2}\left\{x,L_z\right\}+\frac{1}{r_0^2}y\left(-\frac{i}{2}+xp_x+yp_y\right)\nonumber\\
&=&\frac{1}{2r_0^2}\left\{x,L_z\right\}+\frac{1}{r_0^2}y\left(\frac{i}{2}+p_x
x +p_y
y\right)\nonumber \\
&=&\frac{1}{2r_0^2}\left\{x,L_z\right\}+\frac{1}{r_0^2}y\phi_{3,W}.
\earr These expression are not Hermitian if we set $xp_x+yp_y=0$
or $p_x x + p_y y=0$. On the contrary they are Hermitian if we set
$\phi_{3,W}=0$. Therefore the hermiticity of these operators and
their Weyl ordering are strictly correlated. We shall come back to
this remarkable point in the following.

We can now build up any quantity we need in our quantum theory,
for example the $z$ component of the angular momentum
\beq
\label{eq:lz}
L_z=xp_y-yp_x=-i\frac{\partial}{\partial \theta}-\alpha
\eeq
and the Hamiltonian, from (\ref{eq:diarctothamilt})
\barr
\label{eq:Ophamilt2}
H&=&\frac{1}{2}(p_x^2+p_y^2)= \\
&=&\frac{1}{2r_0^2}\left[\left( i\sin\theta\frac{\partial}{\partial
\theta}+\frac{i}{2}\cos\theta+\alpha\sin\theta\right)^2+
\left(-i\cos\theta\frac{\partial}{\partial
\theta}+\frac{i}{2}\sin\theta-\alpha\cos\theta\right)^2\right]\nonumber \\
&=&\frac{1}{2r_0^2}\left(-i\frac{\partial}{\partial
\theta}-\alpha\right)^2+\frac{1}{8r_0^2}. \nonumber \earr One can
check that the ground state energy is
\beq
E_G=\frac{1}{2r_0^2}\left(
\frac{1}{2}-\left|\frac{1}{2}-\overline{\alpha}\right|\right)^2+E_0,
\eeq
where (in ordinary units) $E_0=\frac{\hbar^2}{8mr_0^2}$ and
$\overline{\alpha}\in[0,1[\;$,\quad$\overline{\alpha}=\alpha$ mod
$1$. Notice that we have obtained a constant $E_0$ which was
absent both in the classical Hamiltonians (\ref{eq:classhamiltL0})
and (\ref{eq:standclasshamilt}) and in the quantum Hamiltonian
(\ref{eq:standhamilt}). Let us now discuss the domains in which
these operators are self-adjoint. We see from $L_z$ that a good
domain for its definition is the previously defined
$D_{p_\theta}=\{\psi\in L^2(0,2\pi)|\psi(2\pi)=\psi(0),\psi'\in
L^2(0,2\pi)\}$. In different domains $L_z$ will not be
self-adjoint anymore ad so will not be an observable. The
Schrodinger equation (reinserting the mass and $\hbar$) reads
\beq
i\hbar\frac{\partial \psi}{\partial t}=\frac{\hbar^2}{2mr_0^2}
\left(-i\frac{\partial}{\partial
\theta}-\alpha\right)^2\psi+E_0\psi,
\eeq
which differs from (\ref{eq:standardschreq}) for the presence of
$E_0$.

At this point one should focus on the connection between the
additional term $E_0$ and the quantization procedure we have
established. First of all, observe that if we had not required
that the $p$'s be hermitian we would not have found such a term.
For example, taking $H=L_z^2/2r_0^2$ on the circle in the
classical context and \emph{then} substituting $L_z$ with the
corresponding quantum operator gives (of course) no additional
terms in the Hamiltonian. Another way of getting rid of the
additional terms is by abelian conversion of the algebra
\cite{Kleinert,Faddeev,Batalin}, by introducing an additional set
of coordinates. In practice, there are many ways of dropping or
changing the additional terms without changing the algebra.

Similar terms arise in the quantization on curved manifolds as an
effect of the (intrinsic) curvature of the manifold itself, as
shown by DeWitt \cite{DeWitt} and successively elaborated by
Schulman \cite{Schulman}. We stress however that the constant
$E_0$ found in Dirac's procedure \emph{cannot} be put in direct
correspondence with such curvature terms. Indeed for the
quantization on the circle the scalar curvature is $0$, while the
additional term in Dirac's procedure is $1/8r_0^2$. However, the
fact that no direct proportionality is present between these two
energies does not mean that one of them (namely Dirac's one) is
unphysical, but rather that the procedures of quantization leading
to them reflects different physical processes. While, for example,
in the path-integral procedure no mention of the embedding space
(for example the circle as a subset of the plane) is made, in
Dirac's procedure this embedding is unavoidable. Moreover a
parallel with the classical version of Dirac's procedure hints
that the physical evolution it suggests is like making a small
step (free evolution) in the embedding space followed by a
`projection' on the constraining manifold, obtained by dropping
the component of the step orthogonal to the manifold \cite{Dirac}.
In order to visualize this way of constraining the particle on the
manifold one should follow the procedure we outlined, i.e.\ should
quantize the generators of translations $p_x$ and $p_y$ of the
embedding space and \emph{then} calculate the Hamiltonian. On the
contrary DeWitt's procedures relies upon the presence of a natural
metric on the manifold and the additional term is due to the
intrinsic curvature of the manifold itself. This scenario
obviously sets aside a possible embedding of our manifold in a
bigger one. The choice of a given procedure should depend on the
physical process one has in mind.

Additional work is needed in this direction, for the problem is
certainly far from being solved, as the physical significance of
the additional energies appearing in the different procedures is
not completely understood.

\section{Conclusions}
As we have shown, the Dirac method yields deep insight even in a
simple example like the one we considered. The construction of the
Dirac algebra of brackets is non-trivial and instructive and even
more interesting is the search for an explicit representation of
the self-adjoint operators satisfying the algebra and the
constraints. One must look at their functional form and identify
and interpret any possible freedom inherent to their choice. Then
one must look at their domains of definition, facing sometimes
ordering problems. Eventually, one gains a better comprehension of
the Hamiltonian formalism, the connection between Dirac algebra
and the topology of the constrained manifold and the quantization
procedure on this manifold. An interesting explicit result we have
obtained is the presence of an additional energy term different
from the ones present in other quantization procedures. We have
discussed this term in connection with the Dirac's quantization
procedure arguing that its presence is connected to the (physical)
way of constraining the dynamics on a manifold.

In this paper we have adopted for $p_\theta$ and $L_z$ only the
domain with periodic boundary conditions. Actually there is an
infinity of subsets of $L^2(0,2\pi)$ where every operator we have
considered is self-adjoint, i.e.\ those with
$\psi(2\pi)=e^{i2\pi\beta}\psi(0)$ where $\beta\in[0,1[$. This
issue is clearly exposed in \cite{Gieres,Bonneau} and references
therein. One can regard the gauge transformation with parameter
$\alpha$ in Sec.3.2 as a similarity transformation between these
subsets of $L^2$. The (potential) freedom in the choice of the
domain of definition of the operators is contained in this gauge
transformation.

It would be interesting to elucidate the features of this
formalism, in the form explicitly including the Lagrange
multipliers, in connection with the Faddeev and Popov functional
technique in quantum field theory \cite{Peskin}.

The author would like to thank P.\ Facchi and S.\ Pascazio for
interesting remarks.

\end{document}